# Design and Development of Paper-based Spirometry Device and its Smart-phone based Lung Condition Monitoring and Analysis Software

Yatin Miglani[1], Muddukrishna P[1], Harshvardhan Pande[2], Manoj M. Varma[1*], Bhushan Toley[2*]

1. *Centre for Nanoscience and Engineering*, Indian Institute of Science, Bangalore, India
2. *Department of Chemical Engineering*, Indian Institute of Science, Bangalore, India,
   mvarma@iisc.ac.in, bhushan@iisc.ac.in

*Abstract – We describe the design, construction, and characterization of paper-based devices to perform spirometry, a standard test for lung function assessment. In this device, the instantaneous flowrate of the incoming breath from a person gets transformed to a specific acoustic frequency. In this manner, the time-course of the person's breath profile is mapped to a time-varying acoustic signal. The captured acoustic signal can be converted to standard spirometry curves and relevant parameters can be extracted as we describe in this paper. We compared our device with commercially available devices and showed that these paper-based devices provide similar performance. These devices have the advantage of low-cost, and simplicity of operation compared to currently available commercial products.*

I. INTRODUCTION

According to the data published in The Lancet Respiratory Medicine [1] 545 million people in the world had a chronic respiratory disease in 2017, which was an increase of 39·8 percent from the year 1990. Chronic respiratory diseases accounted for 3·9 million deaths in 2017 (an increase of 18 percent since 1990) and were responsible for 1470 disability-adjusted life-years (DAILYs) per 100,000 individuals (112·3 million total DAILYs, an increase of 13·3 percent since 1990). The most prevalent chronic respiratory diseases were COPD (3·9 percent global prevalence) and asthma (3·6 percent). South Asia had the highest mortality attributable to chronic respiratory disease in the study. For the current scenario, the best possible solution to overcome the problem is the early diagnosis of the disease as these diseases are progressive in nature. Spirometry is a physiological test that plays an important role in the diagnosis and management of COPD (Chronic Obstructive Pulmonary Disease), Asthma, and other lung diseases. Primarily it measures how an individual inspires or expires volume of air as a function of time.

For individuals suffering from severe lung conditions, it is used periodically to monitor their lung condition to provide proper doses of medication. Spirometry test involves a person exhaling air forcefully out of the lungs into the mouthpiece for at least 6 seconds followed by a deep inhalation and measures parameters that are used for the assessment of the lungs. Various spirometry analyses measure lung functioning via. Different parameters, specifically, the amount (volume) and speed (flow-rate) of air that can be inhaled and exhaled. Most clinical spirometers measure a few common parameters, important among which are forced vital capacity (FVC), Forced expiratory volume in the first second of exhalation ($FEV_1$), Peak expiratory flow rate (PEFR), and $FEV_1/FVC$.

For home-based spirometry, currently, there are few devices available commercially. Most of these devices are Peak-flow meters that only measure the Peak expiratory flow rate (PEFR). Although an important parameter, peak-flow rate alone is not a sufficient parameter to diagnose the disorder. This limitation discourages the use of these devices for diagnosis.

Sabharwal et al. [9] made a breakthrough advancement in developing a point-of-care diagnosis device for self-management of asthma using smartphones named *MobileSpiro*. The key contribution of this pressure-sensor-based device was that it detected erroneous patient maneuvers, ensured the quality of the data, and coached the patients with the feedback that could be easily understood. Watanabe et al. [10] have used vortex whistles as flow meters, vortex whistles are low cost and highly portable, and only a subset of common spirometry measurements can be measured reliably. Despite this, modified designs of vortex whistles have been used by various research groups which has given birth to Vortex Spirometry. Larson et al. [11] introduced *Spirosmart*, a smartphone-based spirometer that measures the user's lung function using the phone's built-in microphone. Patel et al. [12] came up with a similar kind of phone-based spirometer called *Spirocall* (*Spirocall* does not require the phone to be a smartphone) which uses a 3D printed vortex whistle with a mobile application to understand a patient's lung condition. They used the whistle audio frequency generated by forced maneuver by the user with time to obtain the parameters. In 2019 Yang et al. [13] Presented smart phone based handheld wireless spirometer that transfers respiratory signal to smart phone using Bluetooth signal transmission for data processing.

Low-cost spirometers ranging in between $10-$50 USD are also available but are capable of only measuring peak expiratory flowrate, these usually use mechanical apparatus with non-eco-friendly one-time use mouthpieces without any electronics, however, are not reliable due to the movements of mechanical parts. Home spirometers in range of $50-$200 USD are also available which report FEV1 but involve either a USB desktop connection or some complex electronics that consume power and expect user to have technical knowledge on how to operate.

In this paper we aimed to 1) Design and development of a low-cost environmentally friendly, zero electronics and home usable paper-based spirometry device capable of producing acoustic signals. 2) Development of frontend Android Application which will record signals and display results. 3) Development of Backend software equipped with machine learning and signal processing algorithms to assist the functioning of whole pulmonary condition analysis system.

## II. DESIGN OF DEVICE

Two devices were developed as a part of this project viz. a square cross-sectional axial device and a flat tangential device.

1. For the square cross-section device, the design contains four parts viz. the outer cover (250 GSM paper), inner rotor support (450 GSM paper), axle (PLA, a biodegradable 3D printing material), rotor (PLA, a biodegradable 3D printing material). The Fig. 1 shows the assembly of the axial device indicating all the parts and blow direction.
2. The working principle of the tangential design is closely related to the siren-based technique. The current design consists of the following parts:
    i. A foldable design of the housing holds the other components inside it.
    ii. Siren disc
    iii. Lower shaft at which siren disc sits.
    iv. Upper shaft around which siren disc rotates.
    v. Vanes are fixed with a siren disc to increase contact between the air stream and the siren disc.

The siren disc has 6 paper made vanes glued to it using Fevibond (synthetic rubber adhesive) with each vane separated by an angle of 60 degrees with the neighboring vanes. The purpose of these vanes is to increase the contact surface of the air stream being blown by the user and the siren disc which would help in better rotation of the shaft. Fig. 1 shows the steps for the assembly of the tangential device.

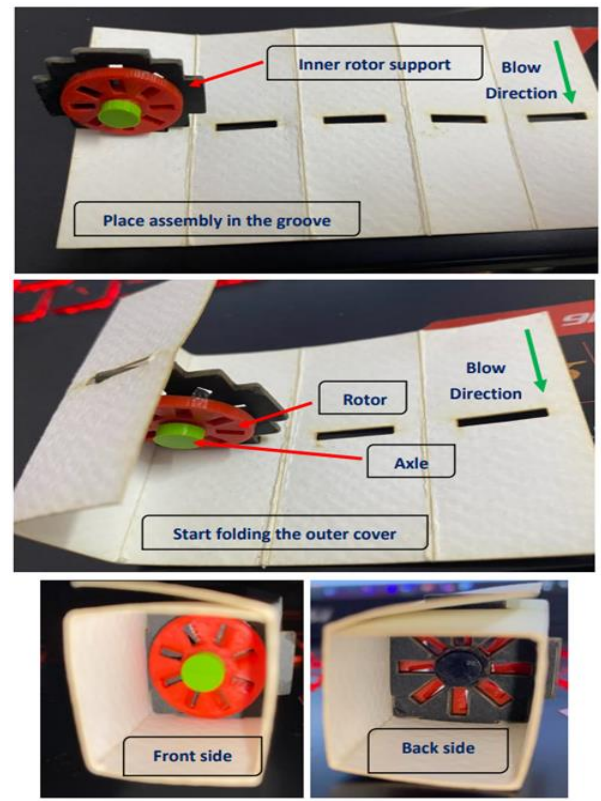

(a) Axial Device (square cross-section)

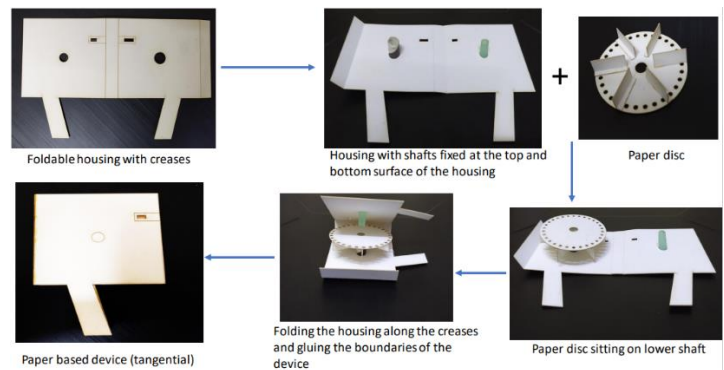

(b) Tangential Device (flat design)

Fig. 1 (a) Axial and (b) Tangential device

III. SIMULATION DETAILS

The Axial device under its operation gets to two states viz. C state and R state, which signify to the sound generation mechanism as C and R refers to contraction and rarefaction respectively. C state appears when the air passage openings overlap among the rotor blades and the inner rotor support. i.e., the air is made to compress as it pass through the rotor blades from the opening of the inner rotor support. R state appears when the air passage opening does not overlap among the rotor blades and the inner rotor support. i.e., the air is made to hit the solid part on the rotor disc as it pass through the opening of the inner rotor support. This generation of contraction and rarefaction in the C and R state of the device eventually becomes the basis for the sound generation. Since the blade openings on the rotor are angled, the air gets a non-zero curl on its way outwards.

Fig. 2 (a) and 2 (b) shows the pressure contours on the outer section of fluid domain of C and R state respectively. From the simulation it can be seen that the outer part of the fluid i.e., air flow will experience high pressure in the case of R state than C state, which in turn leads to higher outward pressure to the outer cover of the device in case of R state than C state, since the R and C state occurs periodically in synchronization of rotor frequency the outer cover eventually goes under periodic stress due to air flowing and thus affects the structural integrity of the outer cover paper. The high pressure in R state appears due to the restricted path of the air molecules, as the air molecules force fully passes through the no overlapping openings the stress in the fluid domain increases, however a region of rarefaction is formed as molecules get farther from each other as can be seen in Fig. 2 (d). Consequently, region of contraction is formed in case of overlapping openings and molecules come close to each other as seen in Fig. 2 (e). Fig. 2 (c) shows the difference between the curl of the air observed as the air exits the device in C and R state respectively, the outward air curl in the C state is higher as the opening on the rotor and inner rotor support overlaps and therefore the air velocity is higher, thus the energy transferred to the outer portion of the fluid domain is lower, thereby the continuous flow keeps the air molecule close to each other hence creating the contraction region. Fig. 2 (d) and 2 (e) shows the air molecule streamlines, the distance between the molecules in C and R state shows the condition of contraction and rarefaction respectively. The shape of the air molecule trajectories continuously fluctuates synchronized with the angular frequency of the rotor, which in turn depends upon the user's blow, and thus becomes the basis of quantization of lung condition. A specific amount of fluctuating curl generates an acoustic signal in space and time of a specific frequency distribution, mathematically,

$$\nabla \times \phi > 0, \nabla_{C\,State} \times \phi > \nabla_{R\,State} \times \phi$$

Flow analysis on Tangential design was performed in order to understand the flow characteristics of the device. The simulation was performed on ANSYS Fluent. These flow simulations helped in the optimization of the design. It became evident from the simulation that most of the airflow was passing through the rear face without being chopped by the holes on the disc shown in Fig. 2 (f) – 2 (g). This suggested modification, where the outlet is changed from the rear face of the device to the holes on the top surface, most of the air passed through the holes on the disc. Simulation is performed when the holes of the disc are perfectly aligned with the holes on the top surface of the housing. The Velocity contour is shown in Fig. 2 (h) – 2 (i). This shows that most of the airflow passes through the first hole found on the disc in the path of airflow. The amount of air flowing through subsequent holes in a clockwise direction decreases. Fig. 2 (g) also shows the airflow from the outlet (holes in the modified design).

For the Axial device it is important to note that the cylindrical part shown in the model image is not the outer cover of the device but the approximated fluid domain for the simulation. The model, flux type and spatial discretization used for the simulation are viscous SST-k-omega, Rhie-chow momentum and second order respectively. The simulation ran for 100 iterations for each time step of 0.01 sec till 5 seconds, and blow inlet velocity of 6 m/s. which is the normal time scale and average human blow velocity for any spirometry test.

Whereas for the Tangential device the flow simulations have been performed using Static Mesh, Viscous turbulent k-epsilon model conditions. For better understanding, dynamic meshing would be required. The purpose of the simulations using Static meshing was to get an idea of whether the velocity streamlines cross the holes on the disc which was essential for the siren technique to work.

IV. FABRICATION METHOD

The Square cross-section device is fabricated using the outer cover (250 GSM paper), inner rotor support (450 GSM paper), axle (PLA, a biodegradable 3D printing material), rotor (PLA, a biodegradable 3D printing material). The paper grooves, shape and openings in inner rotor support are cut using laser cutter, and the whole rotor assembly is assembled to have a working device. Which will produce sound frequency based on the patient's blow maneuver. The Tangential device built here is completely made of paper (GSM-270). AutoCAD 2021 (Autodesk, San Rafael, CA) has been used to design the prototype. 50W $CO_2$ laser in a VLS 3.60 laser engraver (Universal Laser Systems, Scottsdale, AZ) has been used to cut the design on paper. Using a paper of 270 GSM, a rectangular-shaped housing (dimensions: 56mm x 60mm) had been designed (using

AutoCAD 2021) on which both the shafts and the siren disc resides. The housing had been provided with creases on each boundary to fold it and completely close the sides. The design was cut using $CO_2$ laser engraver/cutter.

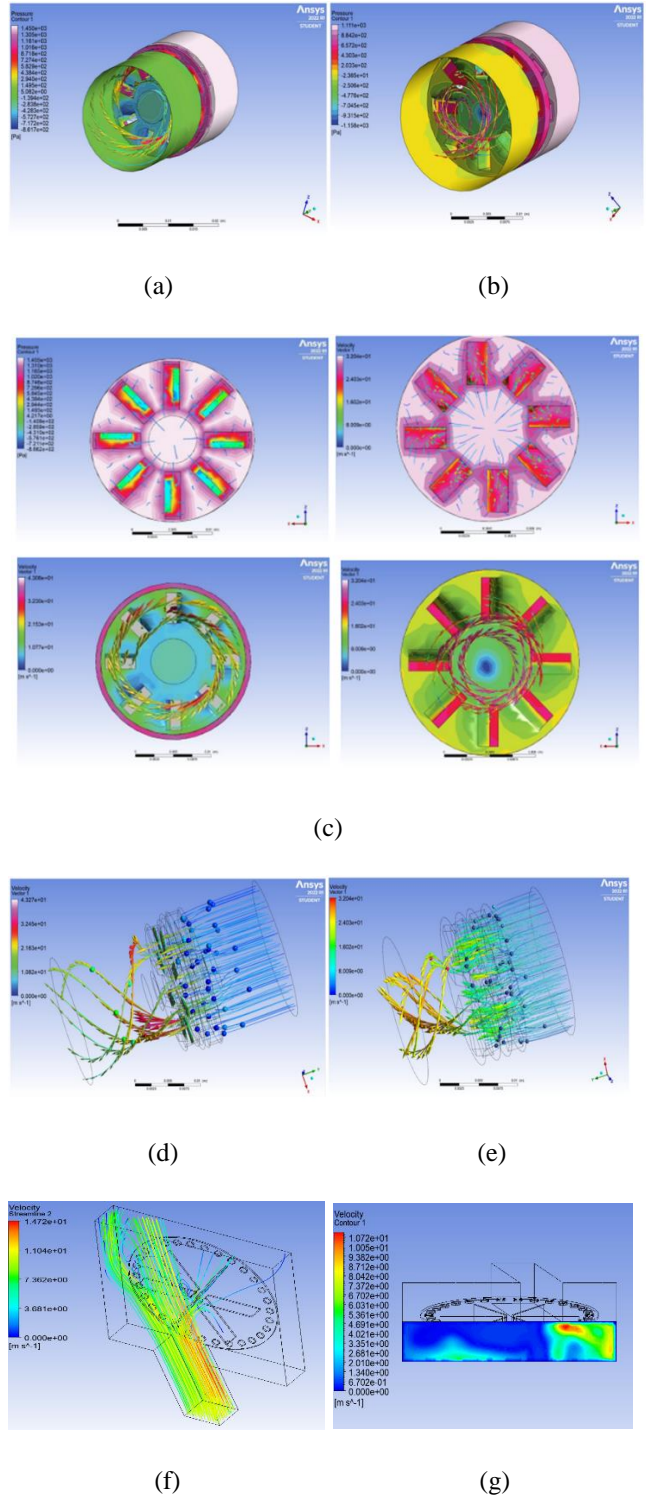

(a)     (b)

(c)

(d)     (e)

(f)     (g)

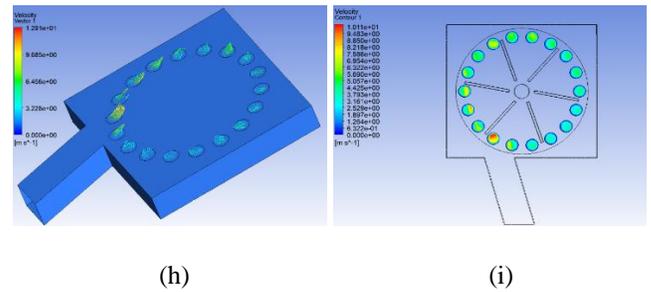

(h)     (i)

Fig. 2 Air flow simulations of Axial (a) - (e) and Tangential devices (f) - (i)

## V. DATA ACQUISITION AND ANALYSIS METHOD

To get a quantifiable measure of the pulmonary condition, a flowrate as a function of time is required, however the developed device produces an acoustic signal, and thus poses a challenge of deriving a dependence of flowrate on time from this acoustic signal. This challenge was solved by the developed algorithm implemented in both front-end android application and as well as back-end analysis software linked to a common server. In a practical sense, the complete system works as follows. Firstly, the patient blows into the device at his/her maximum lung inhale capacity, the device thus produces the acoustic signal which is captured by the microphone of smart-phone having the developed android application. Secondly, this acoustic signal is sent to developed back-end analysis software via common linked server, thus which calculates critical spirometry parameters by extracting and deriving specific signal features and flowrate as a function of time curve from frequency characteristics respectively. Finally, the results of pulmonary condition are sent back to the patient, which the user can see on their smart-phone using the same developed android application.

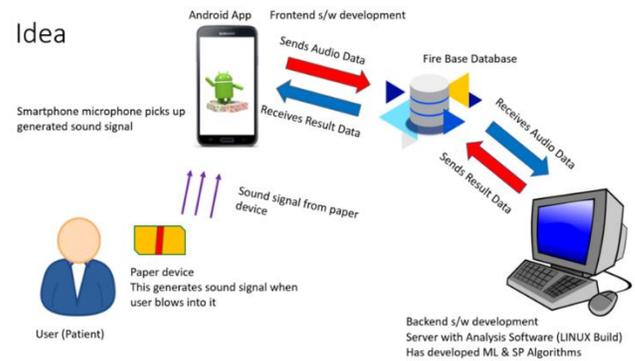

Fig. 3 Idea of the complete Spirometer system

Fig. 4 shows the in line real time processing of the device signal. The backend software is equipped with designed and developed signal processing and machine learning algorithms. The algorithm first finds out the frequency range, distribution, and its dominant components by calculating

Fast Fourier Transform, Power Spectral Density and Spectrogram. Then the algorithm extracts the minimum and maximum frequency present in the signal, algorithm then traces out the points of most dominant frequency using the information of loudness from Spectrogram and Spectral roll off, based on this information algorithm itself designs a filter using the concept of Dynamic Filtering. The designed Dynamic Filter allows a particular band of frequencies to pass through it, this step helps better tracing of the points of dominant frequency and reduces noise in the signal. After the second extraction step the algorithm finds out Frequency vs Time using the minimum value function applied on the first and second curve extraction step averaged out traces. Finally, the algorithm converts the Frequency vs. Time curve to Flow rate vs. Time curve using the relation between the Frequency and Flow rate derived in Machine Learning section. After the conversion to Flow rate vs. Time curve the algorithm calculates the quantifiable spirometry parameters using the Simpsons Method of Integration.

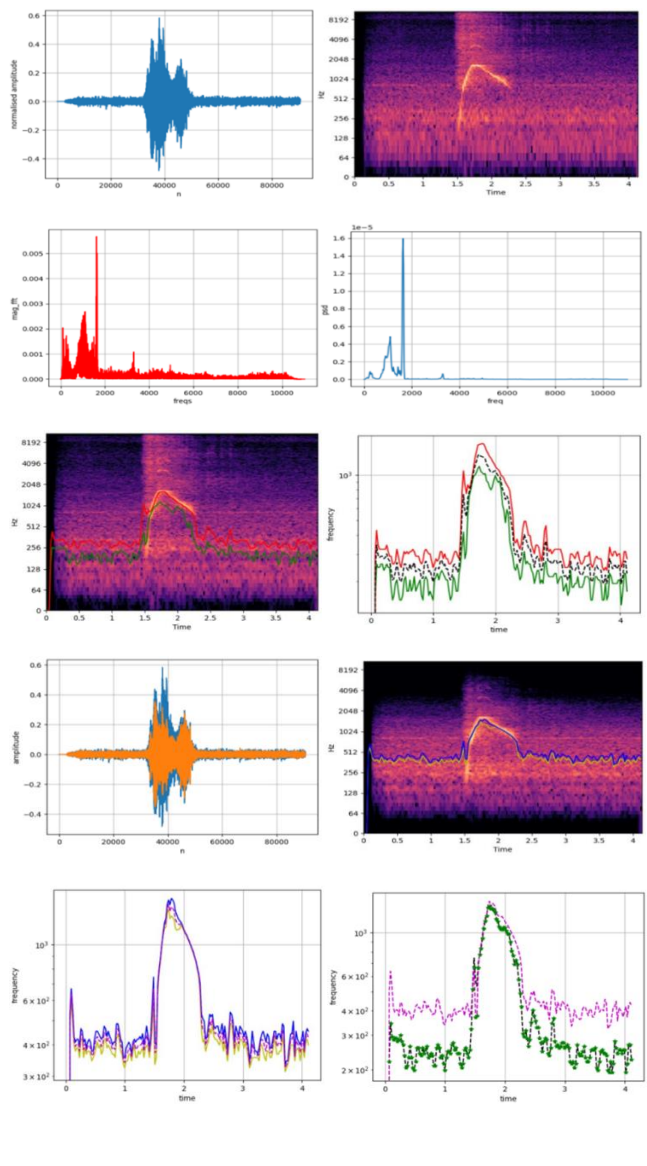

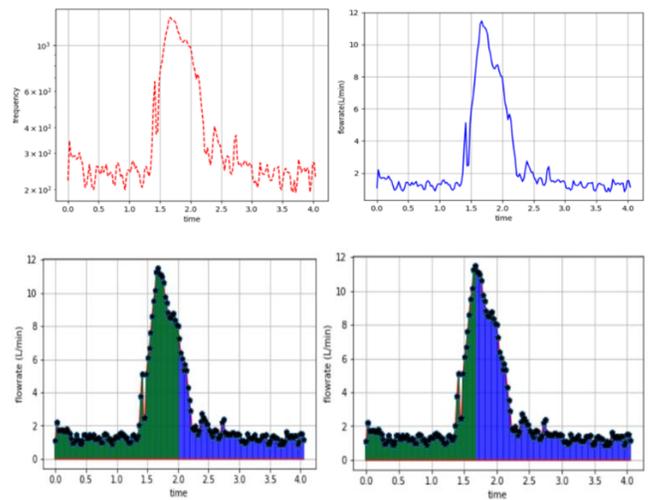

Fig. 4 Signal flow and parameter calculation algorithm.

## VI. CONTROLLED FLOW CHARACTERISATION

To derive the dependence of flowrate on time from the dependence of frequency with time, a converting relationship between flowrate and frequency is required. This relationship between the flowrate and frequency is derived by performing the spirometer test using the developed device and analysis system on constant flow rates, which were produced by the designed blower apparatus. The experiment showed repeatability and relationship between frequency and flowrate was found to be linear.

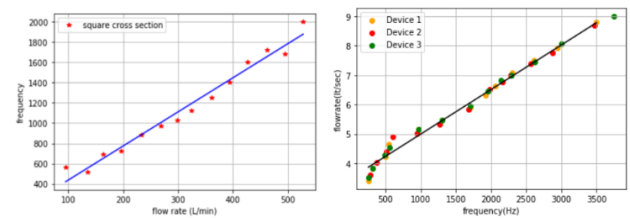

Fig. 5 Flowrate and Frequency dependence Axial device (Left), Tangential device (Right)

Three participants were asked to blow in the device for testing purposes and data collection. The audio of the sound generated from the blow was recorded. The spectrograms obtained from the recorded audio were used to extract the data points for the frequency with time. Using the relation obtained above between frequency and flow rate, Flow-rate vs time was obtained for the extracted data points from the spectrogram. Fig. 5 shows the flow rate vs time curve obtained from spectrogram of audio signal.

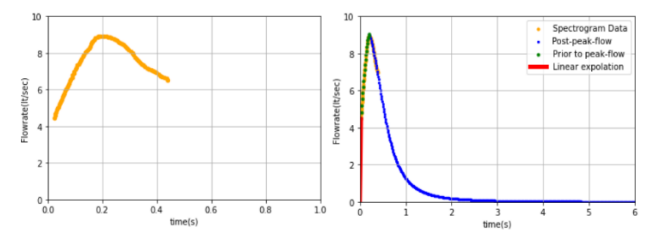

Fig. 6 Flow rate vs time curve observed in experiment (Left) and its extrapolation (Right)

Generally, while undergoing spirometry analysis, patients are asked to forcefully blow for about 6 seconds. Here, while using the paper-based device, the participants were asked to blow to their full extent (known as forced maneuver or forced

Expiration). At least three trials were recorded for each participant and the trial with highest peak frequency was taken for the analysis. Although the blow continued to at least 4 to 5 seconds, the frequency that was tracked via. Spectrogram was around or less than a second. Beyond this point, the sound becomes feeble and the frequency signal gets mixed with interfering background noises. From the curve that had been extracted, the extrapolation technique was used in-order to obtain necessary lung functioning parameters? The function used to extrapolate the flow rate vs time curve was obtained by fitting the curve to the data obtained for flow rate vs time. The curve-fitting technique was used in two separate parts of the curve, prior to reaching the peak flow rate and post peak flow rate. The functions obtained from curve fitting were used for extrapolation of the curve as well as integrating area under the curve.

For the portion of the curve after reaching the peak flow-rate, an inverse sigmoid-like shape was observed in almost all the curves. Inverse Hill-function (a decaying form of Hill-function) was used to fit the portion of the curve after the peak flow rate.

$$F = c \cdot \frac{1 - t^a}{(t^a - b^a)}$$

Where, 't' is time (s), 'c' is peak flow rate, 'F' is flow rate (L/s), 'a' is steepness of the curve and 'b' is time at which flow rate half the peak flow rate.

For the portion of the curve before reaching the peak-flow rate, a cubic function was used to fit the data.

$$F = at^3 + bt^2 + ct + d$$

Where, 't' is time (s), 'F' is flow rate (L/s)

For all the remaining portion of the curve, which is from the origin to the minimum flow-rate data point extracted, was extrapolated linearly. Fig. 6 shows the final flow rate vs time curve used for the estimation of the parameters.

VII. COMPARISON WITH COMMERCIAL DEVICES

For the reliability of the developed device, performance comparison tests were done with two of the commercially available devices (1) Electronic sensor-based respirator and (2) Spirometer Helios 401. Both devices are clinically used for the pulmonary diagnosis. For the comparison test with (1) 3 tests were performed in which simultaneous blow to both respirator and our developed device was given using a 3D printed 'Y' splitter and measurement was done. The readings for the respirator were taken from its commercial software and for the developed device, the developed analysis software was used.

TABLE I. AXIAL DEVICE COMPARISON

| Test | Parameter | Developed device | Respiratory device | %Error |
|---|---|---|---|---|
| 1 | FVC$_{50\%}$ / FVC | 0.356 | 0.382 | 6.84 |
| 1 | FVC$_{75\%}$ / FVC | 0.660 | 0.690 | 4.34 |
| 2 | FVC$_{50\%}$ / FVC | 0.395 | 0.360 | 9.62 |
| 2 | FVC$_{75\%}$ / FVC | 0.736 | 0.777 | 5.27 |
| 3 | FVC$_{50\%}$ / FVC | 0.298 | 0.317 | 5.99 |
| 3 | FVC$_{75\%}$ / FVC | 0.518 | 0.502 | 3.18 |

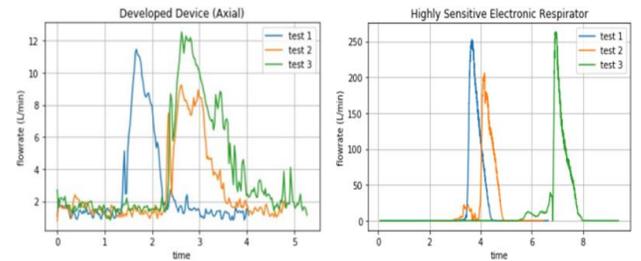

Fig. 7 Comparison table and tests for square cross-section axial device with Electronic sensor based respirator

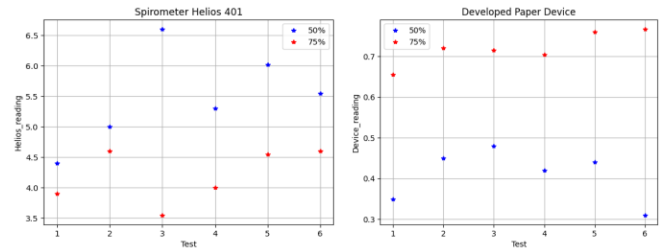

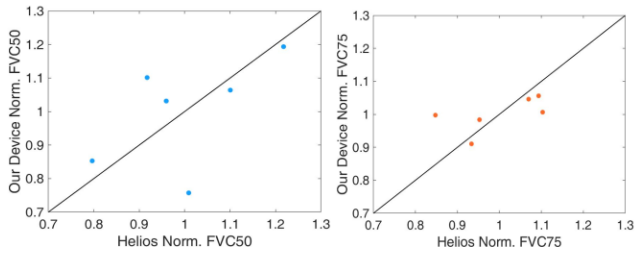

Fig. 8 Comparison test for square cross-section axial device with Helios 401 spirometer.

From the values measured data it is evident that developed device and commercially available electronic respirator device are generating comparable parametric values for the 3 tests.

For the comparison test between spirometer Helios 401 and our developed device 6 independent blow tests were conducted, finally spirometry parameters were measured between the two devices.

First observation between the curves for parameters shown in Figure 7 indicates that for both 50% and 75% the trend of increment and decrement follows, i.e., as the 50% increases or decreases 75% increases or decreases for both the devices. Second observation indicates that a peculiar outlier behavior is observed for both the devices viz at test 3 for Helios 401 and test 6 for the Developed square cross section device. Third observation indicates that for the Helios 401 the values recorded for 50% and 75%, the values for 50% are higher than that of 75%. However, inverse is true for the developed device that is values of 75% are greater than that of 50%. One possible reason for this inverse relation between the two devices is due to fact that Helios 401 gives final output curve as Flowrate vs. Volume on which the further calculations are run, whereas the developed device gives the final output curve as Flowrate vs. Time on which the further calculations are run.

For the comparison test of Tangential device. 3 participants, who did not have a history of any respiratory disorder were asked to perform forced expiration using a paper-based device. The four most important parameters, Forced Vital Capacity (FVC), Forced Expiratory Volume in the first second (FEV1), FEV1/FVC ratio, and Peak-expiratory flow rate (PEFR) were calculated for the participants with the help of function obtained from curve fitting of data points. Fig. 9 (a), 9 (b) and 9 (c) show the spirometry curves predicted for Participants 1, 2 and 3 respectively from the device.

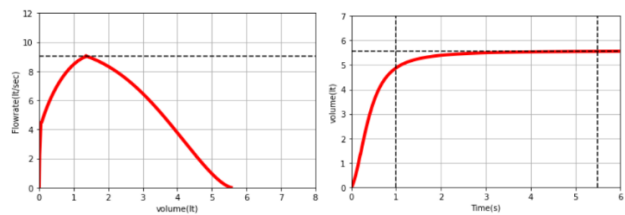

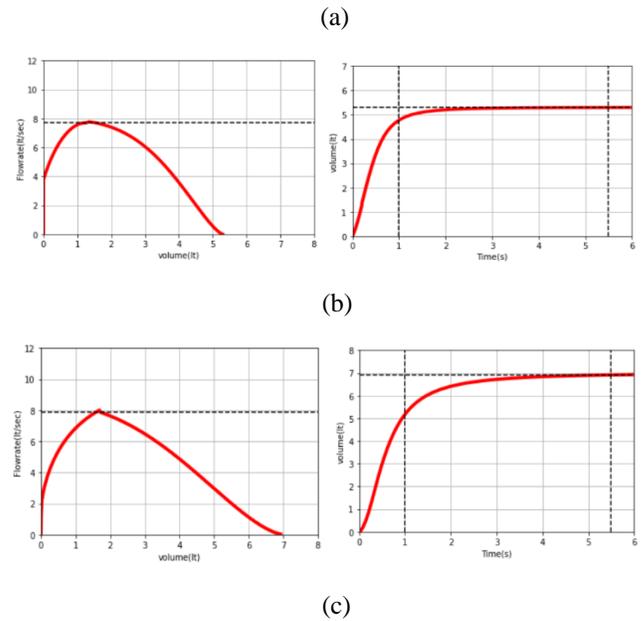

Fig. 9 Spirometry curves predicted for Participants 1, 2 and 3 in plots 9 (a), 9 (b), 9 (c) respectively from the device.

TABLE II. SPIROMETRY PARAMETERS OBTAINED FOR PARTICIPANTS

| ID | FEV1(L) | FVC(L) | FEV1/FVC | PEFR(L/s) |
|----|---------|--------|----------|-----------|
| 1  | 4.83    | 5.45   | 0.88     | 7.79      |
| 2  | 4.91    | 5.6    | 0.87     | 8.8       |
| 3  | 5.2     | 6.87   | 0.76     | 7.9       |

The value has been compared with reference values. The figure shows the reference values commonly used for spirometry analysis. The values obtained for the participants lie in the normal range. From the viewpoint of clinical importance, FEV1/FVC ratio is considered the most important and reliable parameter to decide whether the patient has any abnormality in respiration through spirometry. For normal individuals, the value FEV1/FVC ratio usually lies in the range of 0.70 to 0.9, below which it is considered to be obstruction. A value of FEV1/FVC ratio above 0.9 usually indicates the presence of obstructive disorder. The other parameters like FVC, FEV1, and peak expiratory flow rate, depend upon the reference group of the patient. These reference values depend on the age, height, gender, and race of an individual. Although for the male participant, the values for FVC and FEV1 have been overestimated compared to the normal values, the reference values also vary with the height of an individual. We were not able to acquire data for the reference values according to the height.

A comparison study of the device was also performed with the standard spirometer (RMS Helios 401 Spirometer) for an individual with no history of respiratory disorders. The flow-volume loop and volume-time curve of both the devices (paper-based tangential device and Standard spirometer) are shown. The lung functioning parameters were also compared for both devices. The values of the parameters were found to be under the limits of acceptable error. Fig. 10 shows comparison between the standard spirometer (RMS Helios 401 Spirometer) and the tangential device.

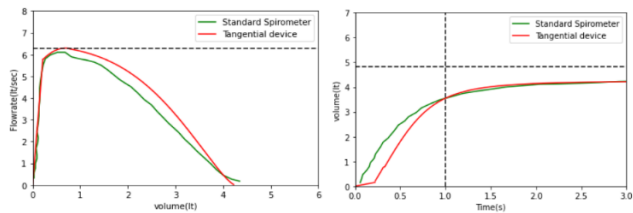

Fig. 10 Comparison between the standard spirometer and the tangential device.

TABLE III. COMPARISON OF LUNG-FUNCTIONING PARAMETERS

| Parameter | Standard Spirometer | Tangential device |
|---|---|---|
| FVC (L) | 4.4 | 4.18 |
| FEV1 (L) | 3.67 | 3.55 |
| FEV1/FVC | 0.834 | 0.85 |
| PEFR (L/s) | 5.96 | 6.257 |

## VIII. ALGORITHM OVERVIEW

Two of the most important algorithms that are deployed in the analysis software are viz (i) Dynamic Filter Design and (ii) Feature Extraction Algorithm.

Implementation of Dynamic Filter Design helps in two main aspects (a) noise reduction in the signal and (b) better feature extraction from the filtered signal. Dynamic Filter Design algorithm generates a transfer function corresponding to a second order Butterworth band pass filter using the signal characteristic data from signal spectrogram, signal FFT and signal PSD Fig. 11 (a) shows the block flow diagram of the Dynamic Filter Design Algorithm.

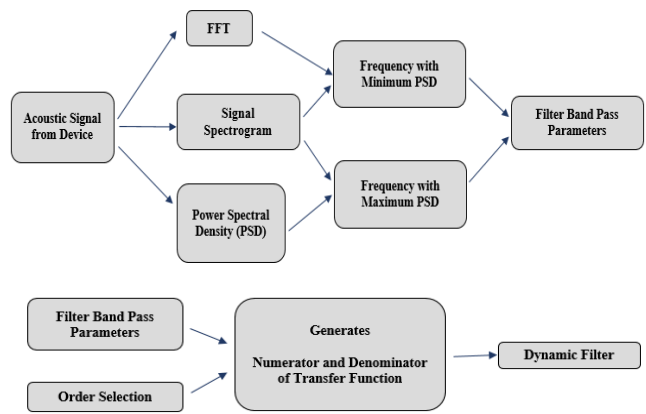

(a) Dynamic Filter Design Algorithm Flow

Implementation of Feature Extraction Algorithm along with Dynamic Filter Design Algorithm traces out the spectrogram lines of the user's correlated acoustic signal with reduced noise levels. Fig. 11 (b) shows the block flow for the Feature Extraction Algorithm.

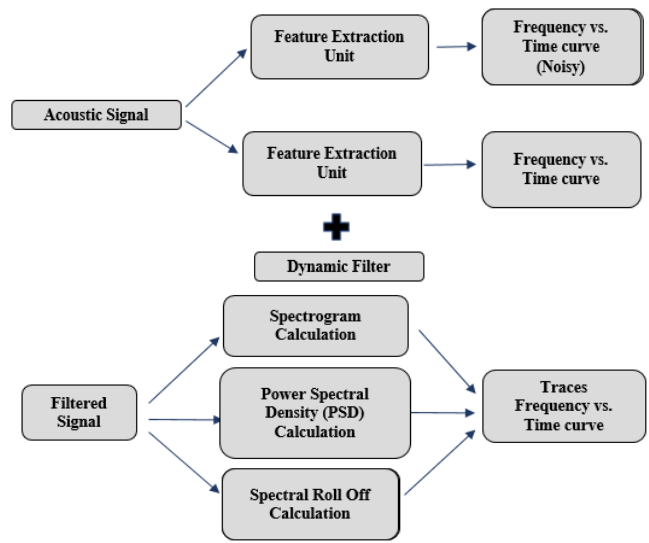

(b) Feature Extraction Algorithm Flow

Fig. 11 Algorithm Flow (a) Dynamic Filter Design Algorithm Flow (b) Feature Extraction Algorithm Flow

## IX. CONCLUSION

A cost effective, home use, user friendly, zero electronic paper-based spirometer device was successfully developed along with its supporting android application and backend software capable of quantifying lung condition. Device shows similar trend behavior in comparison to the commercially available electronic respirator and spirometer (Helios 401). The developed device is not meant to mimic the existing devices but is a new standard itself defining its own independent decision boundaries based on the output values of quantized lung parameters. Following same trend behavior as of complex commercially available devices, the developed device without electronics for sure carries a significant value.

X. ACKNOWLEDGMENT

We thank Dr. Ramakanteswara Beesetty from Novartis, India for his inputs and insights throughout the course of this project.